**From Curriculum Guidelines to Learning Objectives: A Survey of Five Statistics Programs**

Beth Chance and Roxy Peck

Cal Poly – San Luis Obispo

San Luis Obispo, CA 93401

bchance@calpoly.edu

Beth Chance is Professor of Statistics, Cal Poly – San Luis Obispo, CA (e-mail: bchance@calpoly.edu), and Roxy Peck is Professor Emerita of Statistics, Cal Poly – San Luis Obispo, CA. This paper was prepared as part of the ASA workgroup group on revising the Curriculum Guidelines for Undergraduate Programs in Statistical Science. Thanks especially to Nick Horton, Scott Grimshaw, Chris Malone, and other members of the workgroup, Jeff Witmer, John Holcomb, Ginger Rowell, and Cate Rowen for helpful comments on early drafts of the paper.

**ABSTRACT**

The 2000 ASA Guidelines for Undergraduate Statistics majors aimed to provide guidance to programs with undergraduate degrees in statistics as to the content and skills that statistics majors should be learning. With new guidelines forthcoming, it is important to help programs develop an assessment cycle of evaluation. How do we know the students are learning what we want them to learn? How do we improve the program over time? The first step in this process is to translate the broader Guidelines into institution-specific measurable learning outcomes. This paper provides examples of how five programs did so for the 2000 Guidelines. We hope they serve as illustrative examples for programs moving forward with the new guidelines.

## 1. INTRODUCTION

With the growing focus on assessment and accountability, programs at many universities are now expected to define learning outcomes at the program level (as opposed to the course level) and to devise strategies for assessing whether they are being met. In this paper, we will briefly discuss some background definitions related to learning outcomes, suggest strategies for writing learning outcomes, provide examples of undergraduate programs in statistics that developed learning outcomes based on the 2000 ASA Guidelines for Undergraduate Statistics Majors, and discuss possible assessment strategies. It is important to remember that the guidelines provide a high level of autonomy and individuality for different programs. We will try to highlight some of that diversity, while stressing the importance of transparency and accountability.

## 2. THE 2000 ASA GUIDELINES FOR UNDERGRADUATE STATISTICS MAJORS

The 2000 Guidelines were the first effort to provide a coherent description of what it means to have an undergraduate degree in statistics. While undergraduate statistics courses have seen tremendous growth in the last two decades, these guidelines focus on the skills needed specifically by students majoring in statistics. The guidelines exemplify a broad range of skills beyond knowledge of statistical topics, including mathematical, computational, and nonmathematical skills. Suggestions were also given for approaches to teaching these skills, including providing authentic experiences with data and computing and a focus on synthesis and communication. The guidelines were written with the goal of institutional flexibility depending on resources and desired emphases (http://www.amstat.org/education/curriculumguidelines.cfm).

A primary goal of these guidelines was to move beyond "what courses does a statistics major take" to "what can a statistics major *do*" and what skills are needed to carry out these tasks:

> Two alternative approaches to formulating curriculum guidelines emerged: one considered core skills that students should obtain as part of an undergraduate curriculum in statistical science, and the other considered topics that should be included in any such curriculum. Skills and topics are complementary, but an advantage of formulating the curriculum in terms of skills is flexibility. (Bryce, Gould, Notz, and Peck, 2001)

Institutions could then design various approaches to equip students with skills in statistical science (mathematical based and non-mathematics based), computational skills, mathematical foundations, and in a substantive area. A key question becomes how to determine whether graduates of a program are developing the identified skills.

## 3. LEARNING OUTCOMES

### 3.1 What are Learning Outcomes?

Recent efforts to improve higher education have focused on beginning with the development of a mission statement at the level of the academic department (which is then linked to and complementary to university level missions and goals). These program-level mission statements convey the values, philosophy, and vision for a program, highlighting how students will be served and describing the academic environment. Such a student-centered mission statement can then be translated to observable knowledge, skills, abilities, attitudes, and dispositions that should be possessed by graduates of the program. The idea of learning outcomes (also known as learning goals, learning objectives; see Kennedy, 2007 for some history of definitions and distinctions in terminology) has evolved over the last few decades. Kennedy suggests the following working definition:

> Learning outcomes are statements of what a student is expected to know, understand and/or be able to demonstrate after completion of a process of learning.

Wright (2004) claims that defining the key learning goals for the major is the essential first step in the assessment process:

> Goals should focus on what is truly important, not just what can be readily measured, and include skills and attitudes as well as knowledge. Goals should be expressed in three dimensions: 1) *what* we want students to learn; 2) *how well* we want them to perform; and 3) their development *over time*. Goals should represent the highest ambitions of the program, not just minimum standards.

These *outcomes-based* goals, stated in terms of student learning, help characterize graduates of the program much more clearly than a list of required courses.

It is important to remember that program learning outcomes are distinct from a collection of specific content-related goals. Rather, they need to focus on the bigger picture – what are students (as a whole) learning and now able to do as they complete the major. In other words, what is the impact of the program on student learning?

**3.2 How Does One Write Good Learning Outcomes?**

Identification of the important educational goals, such as student learning objectives for the program as well as skills and attitudes appropriate to the major, can be quite challenging and necessitates whole department participation.

The most effective learning outcomes describe program level goals that can be translated into specific, observable, measurable behaviors that can provide evidence of student learning and growth within the program. Many cite the use of "action verbs" as a key component (e.g., the students will be able to … ). The choice of action verb can help determine the level of cognition required (e.g., Bloom's Taxonomy).  For example, *The student will be able to list common statistical methods* (knowledge of facts, terms) versus *The student will be able to classify and explain common statistical methods* (comprehension) versus *The student will be able to apply statistical methods to real-world problems* (application). Program level outcomes that use ambiguous verbs such as *understand*, *know, learn, be aware of, be exposed to, appreciate* should be avoided as they are less easily understood and more difficult to assess. Clearly and succinctly stating what such behaviors will look like, and striving to state the learning outcomes as action-oriented and student-centered, will help in selecting appropriate assessment methods for monitoring the effectiveness of the program.

In looking at the collection of learning outcomes, it is strongly recommended to focus on a small number of important (and realistic) program-level outcomes rather than a large number of superficial course-specific ones.  Some suggest aiming for 5-10 outcomes at the program level. In the Section 4, we profile five institutions who published learning outcomes based on the 2000 Guidelines.

**3.3 Curriculum Mapping**

An important next step after developing learning outcomes is to identify where in the curriculum (e.g., which courses, experiences) students will be developing and practicing the skills necessary to achieve the outcomes. This may consist of a grid identifying the required courses in the program crossed with the learning outcomes.  The grid can also identify whether this is the first exposure to the learning outcome (introductory), a revisit practicing the skills (development), or where students are expected to fully understand and be able to demonstrate mastery of the learning outcome.  A program may want to track student development on the learning outcomes through these stages. The mapping can also indicate the level of feedback provided to students at each stage. An example curriculum mapping is given in Section IV.  Such a curriculum mapping is helpful for identifying gaps in the program (is the curriculum offering the opportunities claimed?) as well as tracking proposed student growth.

## 3.4 Developing an Assessment Plan

The remaining task is development of an ongoing assessment cycle for evaluation and review of the learning outcomes. The effort to develop learning outcomes is wasted if there is not a feedback loop in place, which includes reflection on the results to "close the gap" and improve the program. Programs need a plan for collecting data on student performance and then reflecting, as a department, on the implications for curriculum and pedagogy. The assessments can be direct (e.g., tests, projects) or indirect (e.g., surveys, focus groups). For higher order thinking skills, assessments should be authentic, open-ended, and complex.  Good assessment tasks will capture students' growth over several years.

> … assessment at the program level encompasses much more than grading individual students on a course by course basis.  If a major is viewed as a collection of courses, each with its own content specific objectives, assessment at the program level asks the hard question—what does it all add up to?  What is it that we are hoping this particular collection of courses will produce in the end?  Does it add up to more than just the sum of its content parts?  (Peck and Chance, 2007)

Not every learning outcome has to be assessed every year. A more manageable approach is for programs to identify 2-3 outcomes each year and then cycle through them in a 3-6 year time period. Artifacts can be selected and evaluated outside the regular classroom assessment and rubrics developed (again, ideally a department-wide effect) and applied. The assessment program may identify certain benchmarks for achievement as well (e.g., 70% of students provide evidence of mastery of a particular topic after year three).

## 4. EXAMPLES

Below we focus on five institutions, varying from small undergraduate-only programs to larger Research I universities. The complete learning outcomes published by each institution highlighted here are provided in the appendix. (The details in the appendix are from each institution's department webpage. Some institutions also include more specific performance criteria as well, further clarifying exactly what the student needs to be able to do to meet the learning objective. Information on number of degrees granted in 2013 from http://www.amstat.org/misc/StatsBachelors2003-2013.pdf)

- *Brigham Young University*

    BYU is a private university in Utah with over 30,000 undergraduate students (and 3,000 graduate students). Thirty-five bachelor of science (BS) degrees in statistics and actuarial science were awarded in 2013.

- *Cal Poly – San Luis Obispo*

Cal Poly (CPSLO) is a public primarily undergraduate university in California with close to 20,000 undergraduates. Thirteen BS degrees in statistics were awarded in 2013.

- *North Carolina State University*

NCSU is a public university in Raleigh, NC, with around 35,000 students. The department graduated 21majors in 2013.

- *UC Berkeley*

UC Berkeley (UCB) is a public research university in California with close to 28,000 undergraduate students. Like the other schools in this list, UC Berkeley has begun to see exponential growth in the number of statistics majors since 2010, granting 145 BA degrees in statistics in 2013.

- *Winona State University*

WSU is a comprehensive public university with close to 9,000 students. Roughly 10-15 statistics majors graduate per year.

The table below highlights how some of these programs have translated a curriculum guideline into a learning outcome. For example, whereas the guidelines may call for "background in the following area" and provide a list of topics, these programs have identified more concrete learning outcomes that can be assessed.

Table 1. Program Learning Outcomes

| ASA Curriculum Guideline | Program Learning Outcome |
| --- | --- |
| Graduates should have training and experience in statistical reasoning, in designing studies (including practical aspects), in exploratory analysis of data by graphical and other means, and in a variety of formal inference procedures. | Be able to synthesize and apply knowledge of common inferential methods, understanding the limitations of the procedures and the appropriate scope of conclusions (CPSLO) |
| Design of studies (e.g., random assignment, replication, blocking, analysis of variance, fixed and random effects, diagnostics in experiments; random sampling, stratification in sample surveys; data exploration in observational studies) | Demonstrate good working knowledge of … efficient design of studies (WSU) |
| Programs should require familiarity with a standard statistical software package and encourage study of data management and algorithmic problem solving. | Demonstrate competence in database concepts and terminology through preparation for the SAS Certified Base and Advanced Programmer exams (BYU) |
| Graduates should be expected to write clearly, speak fluently, and have developed skills in collaboration and teamwork and organizing and managing projects. | Explain statistical ideas, methods and results orally and in writing to non-statistical audiences. (NCSU) |
| Undergraduate major programs should include study of probability and statistical theory, along with the prerequisite mathematics, especially calculus and linear algebra. | Solve probability problems in finite sample spaces, with discrete and continuous univariate random variables, and apply the Central Limit Theorem (BYU) |

Presumably some vague ideas such as "good working knowledge" are more clearly defined internally. For example, at Cal Poly, the sixth learning outcome (row 1 in table above) is broken down into these following observable behaviors:

- Select appropriate inferential procedures for question at hand based on research question of interest, types, and number of variables, power considerations, etc.
- Understand that nonparametric alternatives exist for most parametric procedures, and recognize advantages and limitations of such procedures.
- Identify and explain limitations of procedures such as extrapolation and overfitting; lack of significance is not the same as evidence of no effect and statistical significance does not imply practical significance or importance.
- Use and interpret analysis results, such as interpreting model coefficients and making predictions.
- State appropriate context- and design-specific conclusions from an analysis.

Next a program can identify the courses where the learning outcomes are addressed. Table 2 highlights a small portion of the curriculum mapping at BYU for two of the learning outcomes.

Table 2. Example Curriculum Mapping

|  | Experimental Design and Analysis | Communication |
|---|---|---|
| Stat 121: Principles of Statistics | Preparatory | Preparatory |
| Stat 230: Analysis of Variance | Introductory | Introductory |
| Stat 290: Communication of Statistical Results |  | Intermediate |
| Stat 431: Experimental Design | Advanced |  |
| Stat 497R: Introduction to Statistical Research |  | Advanced |

The curriculum mapping may include support and GE courses as well as required courses in the major. The map is helpful for ensuring that students are given sufficient opportunity to practice and master a learning outcome, and also for a program to identify appropriate opportunities for assessment.

For example, Cal Poly has been giving the same assessment questions during a two-course introductory sequence and at the beginning of the senior-level capstone course in an attempt to capture student gains in student understanding and performance. These questions have been helpful in identifying weaknesses by our graduating seniors on key statistical concepts, helping to identify areas of weakness in the overall curriculum, leading to departmental discussion on how to address these deficiencies. For example, Cal Poly noted that many of their students equate ANOVA with experimental design and Regression with observations studies. This may be explained by the fact that the regression modelling course is separate and not in sequence with the experimental design course. This has led them to consciously incorporate more experimental data sets into the Regression course. In addition, they have collected data on embedded exam questions in different courses. Again, the goal is to evaluate the program, separately from evaluating individual students or teaching effectiveness.

The capstone course also provides Cal Poly the opportunity to evaluate student work on more involved projects. For example, faculty jointly developed a computing assessment that was then evaluated for choice of technology tools as well as documentation and communication of results. In addition, students complete a senior project – an independent research or consulting project over two quarters. Cal Poly recently instituted oral presentations by the students which are evaluated by faculty audience members. They have developed a rubric for evaluating these presentations which are now utilized for student presentations in the introductory courses for our majors as well.

Additionally, Cal Poly administers an online "exit survey" of graduating seniors, asking their input on the curriculum as a whole, on specific courses, and on the utility of the opportunities they were provided. Cal Poly has now collected data for over ten years, and can compare how student performance has changed and fluctuated over time and as different changes to the curriculum are implemented. Though they may only evaluate a few students each year, combing the data over the years helps to identify consistent trends. They can also space out different assessments over time so that they aren't using the same ones every year but still eventually acquire comparison data within the institution.

Contact the first author if you want to see examples of these items and rubrics.

## 5. OPEN QUESTIONS

In moving forward to update the 2000 ASA guidelines, and in particular to enable use of these guidelines to evaluate programs, institutions may be well served by operationalizing more of the definitions. For example, what does" have good working knowledge" mean? Should the "fundamentals" and "most commonly used methods" be explicitly stated and defined similarly for all programs? As a community do we want to aim toward a more uniform approach to developing and assessing communication skills within and across disciplines? How should skills such as team work and statistical consulting be evaluated? Should the guidelines include specific mention of ethics in statistical practice? Are there any of these decisions where they ASA would like more influence or should all of this be left up to the individual institutions?

In moving forward with the revision of the undergraduate curriculum guidelines, the ASA could serve a very useful role in providing assessment resources and feedback to programs as more the "skills and topics" suggested in the guidelines are translated into observable student behavior. The new guidelines may wish to clarify or further operationalize some definitions, while leaving a high level of autonomy and individuality for different programs. The ASA may also wish to help institutions to develop and coordinate assessment efforts of these guidelines.

# APPENDIX: PROFILE OF UNIVERSITIES SURVEYED

*Brigham Young University* (https://learningoutcomes.byu.edu/#college=-Z6czUKBNzwp&department=ra8j4G2EhKRn&program=iTxl2sGUjlCW)

Students graduating with a BYU BS Statistics degree will demonstrate competence in the statistical, mathematical, computational, and nonmathematical skills in the American Statistical Association Curriculum Guidelines for Undergraduate Programs in Statistical Science. More specifically, they will

1. Statistical Methods

a) Demonstrate the design and analysis of randomized factorial experiments and blocking at the level of a professional statistician

b) Demonstrate multiple regression modeling at the level of a professional statistician

c) Demonstrate competence in fitting logistic regression and fitting ARIMA time series models

2. Statistical theory

a) Solve probability problems in finite sample spaces, with discrete and continuous univariate random variables, and apply the Central Limit Theorem

b) Demonstrate the derivation of frequentist and Bayesian inference for one-sample proportions and means

3. Computing

a) Write a computer program and use professional statistical software for regression and design of experiments

b) Demonstrate competence in database concepts and terminology through preparation for the SAS Certified Base and Advanced Programmer exams

4. Mathematics

a) Apply the results of differential, integral, and multivariate calculus to problems in probability and mathematical statistics

b) Demonstrate competence with matrix computation and apply results from linear algebra to the linear model

5. Communication

a) Write technical reports and make technical presentations containing statistical results, and work in teams to demonstrate the consulting skills of a professional statistician

---

*Cal Poly* (http://statistics.calpoly.edu/content/about-us)

The Statistics Department offers a degree program leading to the B.S. degree in Statistics in the College of Science and Mathematics.

The Statistics Department graduate will:

1. Have good working knowledge of the most commonly used statistical methods, including

a) statistical modeling and the omnipresent role of variability

b) efficient design of studies and construction of effective sampling plans

c) exploratory data analysis

d) formal inference process

2. Have background in probability, statistical theory, and mathematics, including especially calculus, linear algebra and symbolic and abstract thinking

a) Be able to synthesize and apply knowledge of common inferential methods, understanding the limitations of the procedures and the appropriate scope of conclusions

b) Communicate effectively (written and oral) with skills in collaboration (within and between

disciplines) and teamwork, and in organizing and managing projects

c) Have good mastery of several standard statistical software packages and facility with data management strategies

d) Have a focused concentration in an area of application outside the discipline of statistics

The Statistics Department graduate will have received:

• Experience with real data and authentic applications

• Frequent opportunities to develop communication skills

• Capstone experiences for students

• Frequent interaction with faculty and their peers and timely advising

• Exposure to statisticians and statistical applications outside the Cal Poly community

*North Carolina State University* (http://www.stat.ncsu.edu/programs/ugrad/)

Upon completion of our program, students will be able to:

1. Explain the theoretical basis of commonly used statistical methods.
2. Design sample surveys and experiments for standard situations.
3. Correctly analyze and interpret the results from standard designed experiments, sample surveys, and observational studies.
4. Demonstrate sufficient computer programming ability to manage data, implement standard statistical methods, and learn new programming languages in the future.
5. Explain statistical ideas, methods and results orally and in writing to non-statistical audiences.

---

*UC Berkeley* (http://statistics.berkeley.edu/programs/undergrad/learninggoals)

Majors are expected to learn concepts and tools for working with data and have experience in analyzing real data that goes beyond the content of a service course in statistical methods for non-majors. Majors should understand

1. the fundamentals of probability theory
2. statistical reasoning and inferential methods
3. statistical computing
4. statistical modeling and its limitations

and have skill in

5. description, interpretation and exploratory analysis of data by graphical and other means
6. effective communication.

*Winona State University*

The statistics graduate will:

1. Demonstrate good working knowledge of the most commonly used statistics methods, including:

a) exploratory data analysis, basic inference, and limitations of procedures

b) statistical modeling

c) efficient design of studies

2. Demonstrate their knowledge of basic mathematical skills needed for statistics, including:

a) probability and statistical theory

b) calculus foundations

c) symbolic and abstract thinking

d) linear algebra

3. Prepare and present both written and oral presentations of statistical results for various audiences. This involves:

a) effective communication with statisticians

b) effective communication with non-statisticians

4. Implement basic computer science skills needed for statistics, including:

a) data management tools

b) basic programming algorithms and logic

c) use of a statistical software package for standard analyses

5. Demonstrate competency as a statistician by completing and independent data analysis, research project, or internship.